\documentclass[aoas,nameyear,seceqn,dvips]{arximspdf}
\usepackage{mathbh,multirow}
\usepackage{xypic}
\usepackage{graphics}
\doi{10.1214/09-AOAS288}
\volume{4}
\issue{1}
\pubyear{2010}
\firstpage{320}
\lastpage{339}

\makeatletter
\newtheorem{theorem}{Theorem}
\newproclaim{definition}{Definition}
\newproclaim{result}{Result}
\newproclaim{Remarks}{Remarks}
\makeatother

\begin{document}
\begin{frontmatter}

\title{Causal graphical models in systems genetics: A~unified
framework for
joint inference of causal~network and genetic architecture
for~correlated~phenotypes\thanksref{A1}}
\runtitle{Causal graphical models in systems genetics}
\thankstext{A1}{Supported by CNPq Brazil (ECN); NIDDK Grants DK66369,
DK58037 and DK06639
(ADA, MPK, BSY, ECN); and by NIGMS Grants PA02110 and GM069430-01A2
(BSY).}

\begin{aug}
\author[a]{\fnms{Elias} \snm{Chaibub Neto}\ead[label=e1]{chaibub@stat.wisc.edu}},
\author[b]{\fnms{Mark P.} \snm{Keller}\ead[label=e2]{mkeller3@wisc.edu}},
\author[b]{\fnms{Alan D.} \snm{Attie}\ead[label=e3]{adattie@wisc.edu}}\\
\and
\author[a]{\fnms{Brian S.} \snm{Yandell}\ead[label=e4]{byandell@wisc.edu}\corref{}}

\runauthor{Chaibub Neto, Keller, Attie and Yandell}

\affiliation{University of Wisconsin--Madison}

\address[a]{E. Chaibub Neto\\
B. S. Yandell\\
Department of Statistics\\
University of Wisconsin--Madison\\
1300 University Avenue\\
Madison, Wisconsin 53706\\ USA\\
\printead{e1}\\
\phantom{E-mail: }\printead*{e4}}

\address[b]{M. P. Keller\\
A. D. Attie\\
Department of Biochemistry\\
University of Wisconsin--Madison\\
433 Babcock Drive\\
Madison, Wisconsin 53706\\ USA\\
\printead{e2}\\
\phantom{E-mail: }\printead*{e3}}
\end{aug}

\received{\smonth{12} \syear{2008}}
\revised{\smonth{7} \syear{2009}}

%
\begin{abstract}
Causal inference approaches in systems genetics exploit quantitative
trait loci (QTL) genotypes to infer causal relationships among
phenotypes. The~genetic architecture of each phenotype may be
complex, and poorly estimated genetic architectures may compromise
the inference of causal relationships among phenotypes. Existing
methods assume QTLs are known or inferred without regard to the
phenotype network structure. In this paper we develop a QTL-driven
phenotype network method (\mbox{QTLnet}) to jointly infer a causal
phenotype network and associated genetic architecture for sets of
correlated phenotypes. Randomization of alleles during meiosis and
the unidirectional influence of genotype on phenotype allow the
inference of QTLs causal to phenotypes. Causal relationships among
phenotypes can be inferred using these QTL nodes, enabling us to
distinguish among phenotype networks that would otherwise be
distribution equivalent. We jointly model phenotypes and QTLs using
homogeneous conditional Gaussian regression models, and we derive a
graphical criterion for distribution equivalence. We validate the
\mbox{QTLnet} approach in a simulation study. Finally, we illustrate with
simulated data and a real example how \mbox{QTLnet} can be used to infer
both direct and indirect effects of QTLs and phenotypes that co-map
to a genomic region.
\end{abstract}

%

\end{frontmatter}
%

\section{Introduction}

In the past few years it has been recognized that genetics can be
used to establish causal relationships among phenotypes organized in
networks [Schadt et al. (\citeyear{ScLaYaZhirkt2005}), Kulp and Jagalur (\citeyear{KuJa2006}), Li et al. (\citeyear{LiTsShStWePaCh2006}),
Chen, Emmert-Streib and Storey (\citeyear{ChEmSt2007}), Liu, de la Fuente and Hoeschele (\citeyear{LideHo2008}),
Aten et al. (\citeyear{AtFuLu2008}), Chaibub Neto et
al. (\citeyear{ChFeAtYa2008})]. These approaches aim to generate a hypothesis about causal
relationships among phenotypes involved in biological pathways
underlying complex diseases such as diabetes. A key element in these
methods is the identification of quantitative trait loci (QTLs) that
are causal for each phenotype. The genetic architecture of each
phenotype, which consists of the locations and effects of detectable
QTLs, may be complex. Poorly estimated genetic architectures may
compromise the inference of causal relationships among phenotypes.
Existing methods that estimate QTLs from genome scans that ignore
causal phenotypes bias the genetic architecture by incorrectly
inferring QTLs that have indirect effects.

In this paper we propose a novel framework for the joint inference
of phenotype network structure and genetic architecture (\mbox{QTLnet}). We
model phenotypes and QTL genotypes jointly using homogeneous
conditional Gaussian regression (HCGR) models [Lauritzen (\citeyear{La1996})]. The
genetic architecture for each phenotype is inferred conditional on
the phenotype network. Because the phenotype network structure is
itself unknown, the algorithm iterates between updating the network
structure and
genetic architecture using a Markov chain Monte Carlo (MCMC) approach. The
posterior sample of network structures is summarized by Bayesian
model averaging. To the best of our knowledge, no other proposed
method explicitly uses an inferred network structure among
phenotypes when performing QTL mapping. Tailoring QTL mapping to
network structure avoids the false detection of QTLs with indirect
effects and improves phenotype network structure inference.

We employ a causal inference framework with components of both
randomized experiments and conditional probability. Randomization of
alleles during meiosis and the unidirectional influence of genotype
on phenotype allow the inference of causal QTLs for phenotypes.
Causal relationships among phenotypes can be inferred using these
QTL nodes, enabling us to distinguish between networks that would
otherwise be distribution equivalent.

We are particularly interested in inferring causal networks relating
sets of phenotypes mapping to coincident genomic regions. It is
widely asserted that alleged ``hot spots'' may have a ``master
regulator'' and that most co-mapping is due to indirect effects
[Breitling et al. (\citeyear{BrLiTeirkt2008})]. That is, such a hot spot QTL could influence
a single phenotype that is upstream of many others in a causal
network; ignoring the phenotype network would result in a perceived
hot spot. One objective of our \mbox{QTLnet} method is to sort out the
direct and indirect effects of QTLs and phenotypes in such
situations.

In the next sections we develop in detail a framework for the joint
inference of causal network and genetic architecture of correlated
phenotypes. The core idea is to learn the structure of mixed Bayesian
networks composed of phenotypes (continuous variables) and QTLs
(discrete variables). We model the conditional distribution of
phenotypes given the QTLs with homogeneous conditional Gaussian
regression models described in Section \ref{sec2}. This allows us to justify
formal inference of causal direction along the Bayesian networks in
Section \ref{sec3}. That is, we can reduce the size of equivalence classes of
Bayesian networks of phenotypes by using driving QTL. In Section \ref{sec4} we
show how a conditional LOD score can formally measure conditional
dependence among phenotypes and QTLs and how QTL mapping can be
embedded in a graphical models framework. Section \ref{sec5} presents the MCMC
approach for \mbox{QTLnet}. Simulation studies in Section \ref{sec6} validate the
\mbox{QTLnet} approach and provide an explanation for some QTL hot spots.
Section~\ref{sec7} uses real data to illustrate how \mbox{QTLnet} can be used to infer
direct and indirect effects of QTLs and phenotypes that co-map to a
genomic region. The \hyperref[sec8]{Discussion} puts this work in the context of open
questions. Proofs of formal results are given in the Supplement
[Chaibub Neto et al. (\citeyear{ChKeAtYa2009})].

\section{HCGR genetic model}\label{sec2}

In this section we recast the genetical model for QTL studies as a
homogeneous conditional Gaussian regression model that
jointly models phenotypes and QTL genotypes. Conditional on the QTL
genotypes and covariates, the phenotypes are distributed according
to a multivariate normal distribution. The QTLs and covariates enter
the HCGR model through the mean in a similar fashion to the
seemingly unrelated regression model [Banerjee et al. (\citeyear{BaYaYi2008})].
However, the correlation structure among phenotypes is explicitly
modeled according to the directed graph representation of the
phenotype network. We derive the genetic model from a system of
linear regression equations and show that it corresponds to a
homogenous conditional Gaussian regression model.

In QTL studies, the observed data consist of phenotypic trait
values, $\mathbf{y}$, and marker genotypes, $\mathbf{m}$, on $n$
individuals derived
from an inbred line cross. Following Sen and Churchill (\citeyear{SeCh2001}), we
condition on unobserved QTL genotypes, $\mathbf{q}$, to partition our model
into genetic and recombination components, respectively relating
phenotypes to QTLs and QTLs to observed markers across the genome,
\[
p(\mathbf{y}  ,   \mathbf{q}\mid\mathbf{m})   =   p(\mathbf
{y}\mid\mathbf{q}  ,   \mathbf{m})   p(\mathbf{q}
\mid\mathbf{m})   =   p(\mathbf{y}\mid\mathbf{q})   p(\mathbf
{q}\mid\mathbf{m}) ,
\]
where the second equality follows from conditional independence,
$\mathbf{y}
\perp \!\!\!\!\perp \mathbf{m}\mid\mathbf{q}$. That is, given the
QTL genotypes, the marker
genotypes provide no additional information about the phenotypes.
Estimation of the recombination model, $p(\mathbf{q}\mid\mathbf
{m})$, is a
well-solved problem [Broman et al. (\citeyear{BrWuSeCh2003})] and is not addressed in this paper.

Let $i=1,\ldots,n$ and $t=1,\ldots,T$ index individuals and phenotype
traits, respectively. Let $\mathbf{y}=(\mathbf{y}_1,\ldots,\mathbf
{y}_n)'$ represent all
phenotypic trait values, $\mathbf{y}_i = (y_{1i},\ldots,y_{Ti})'$ represent
the measurements of the $T$ phenotype traits for individual $i$, and
let $\bolds{\varepsilon}_i = (\varepsilon_{1i},\ldots,\varepsilon_{Ti})'$
represent the
associated independent normal error terms. We assume that individual
$i$ and trait $t$ have the following phenotype model:
%
\begin{equation}
y_{ti}   =   \mu^{\star}_{ti}   +   \sum_{v \in \operatorname{pa}(y_t)}
\beta_{tv}   y_{vi}   +   \varepsilon_{ti} , \qquad       \varepsilon_{ti}
  \sim  N(0,\sigma^2_t),
\label{phenotypemodel}
\end{equation}
where $\mu^{\star}_{ti} = \mu_t + \mathbf{X}_{ti}   \bolds{\theta
}_t$, where $\mu
_t$ is the overall mean for trait $t$, $\bolds{\theta}_t$ is a column
vector of
all genetic effects constituting the genetic architecture of trait $t$
plus any additional additive or interactive covariates, and $\mathbf{X}_{ti}$
represents the row vector of genetic effects predictors derived from
the QTL genotypes along with any covariates. The notation $\operatorname{pa}(y_t)$
represents the set of parent phenotype nodes of $y_t$, that is, the set
of phenotype nodes that directly affect $y_t$. Genetic effects may
follow Cockerham's genetic model, but need not be restricted to this
form [Zeng, Wang and Zou (\citeyear{ZeWaZo2005})].

The Jacobian transformation from $\bolds{\varepsilon}_i \rightarrow
\mathbf{y}_i$ allows us to
represent the joint density of the phenotype traits conditional on the
respective genetic architectures as multivariate normal with the
following mean vector and covariance matrix.

\begin{result} \label{concmatrixformula}
The conditional joint distribution of the phenotype traits organized
according to the set of structural equations defined in (\ref
{phenotypemodel}) is $\mathbf{y}_i \mid\bolds{\mu}^{\star}_{i},
\bolds{\beta},   \bolds{\sigma}
^2   \sim  N_{T}(\bolds{\Omega}^{-1}   \bolds{\gamma
}_i   ,
\bolds{\Omega}^{-1} )$, where $\bolds{\mu}^{\star}_{i} =
(\mu^{\star
}_{1i},\ldots,\mu^{\star}_{Ti})'$, $\bolds{\beta}= \{ \beta_{tv}
\dvtx  v \in
\operatorname{pa}(y_t),   t = 1, \ldots, T \}$, $\bolds{\sigma}^2 = (\sigma
^2_{1},\ldots
,\sigma^2_{T})'$, $\bolds{\Omega}$ is the concentration
matrix with entries given by
\[
\omega_{tv} = \cases{\displaystyle
\dfrac{1}{\sigma^2_t} + \sum_{s}   \dfrac{\beta^{2}_{st}}{\sigma
^2_s} \mathbh{1}_{\{t \rightarrow s\}}, & \quad for $t=v$,\vspace*{2pt}
\cr
\displaystyle-\dfrac{\beta_{vt}}{\sigma^2_v} \mathbh{1}_{\{t \rightarrow v\}} -
\dfrac{\beta_{tv}}{\sigma^2_t} \mathbh{1}_{\{v \rightarrow t\}} +
\sum_s
\dfrac{\beta_{sv}   \beta_{st}}{\sigma^2_v} \mathbh{1}_{\{v
\rightarrow s
  ,  t \rightarrow s \}}, & \quad for $t \not= v$,}
\]
$\bolds{\gamma}_i$ is a vector with entries
$\frac{\mu^{\star}_{ti}}{\sigma^2_t}   -   \sum_{s \not= t}
\frac{\beta_{st}   \mu^{\star}_{si}}{\sigma^2_s} \mathbh{1}_{\{t
\rightarrow s\}}$, and $\mathbh{1}_{\{t \rightarrow s\}}$ is the indicator
function that trait $t$ affects trait $s$.
\end{result}

\begin{Remarks*}
 (1) The model allows different genetic architectures
for each phenotype. (2) The covariance structure depends exclusively
in the relationships among phenotypes since $\bolds{\Omega}$ depends
only on
the partial regression coefficients relating phenotypes ($\beta$'s)
and variances of error terms ($\sigma^2$'s), and not on the genetic
architectures defined by the $\bolds{\theta}$'s. (3) When the correlation
between two phenotypes arises exclusively because of a pleoitropic
QTL, conditioning on the QTL genotypes makes the phenotypes
independent; thus, the concentration matrix of the conditional model
does not depend on the genetic architecture. (4) This model can
represent acyclic and cyclic networks. However, we focus on acyclic
networks in this paper.
\end{Remarks*}

We now show that our model corresponds to a homogeneous conditional
Gaussian regression model. The conditional Gaussian (CG) parametric
family models the covariation of discrete and continuous random
variables. Continuous random variables conditional on discrete
variables are multivariate normal [Lauritzen (\citeyear{La1996})]. The joint
distribution of the vectors of discrete ($\mathbf{q}_i$) and continuous
($\mathbf{y}_i$) variables have a density $f$ such that
%
\begin{equation}
\log{f(\mathbf{q}_i   ,   \mathbf{y}_i)}   =   g(\mathbf{q}_i)
  +   \mathbf{h}'(\mathbf{q}_i)
\mathbf{y}_i
  -   \mathbf{y}_i'   \mathbf{K}(\mathbf{q}_i)   \mathbf{y}_i / 2,
\end{equation}
where $g(\mathbf{q}_i)$ is a scalar, $\mathbf{h}(\mathbf{q}_i)$ is a
vector and $\mathbf{K}(\mathbf{q}_i)$
is a positive definite matrix. The density $f$ depends on observed
markers $\mathbf{m}_i$ as
%
\begin{equation}
\log{f(\mathbf{q}_i   ,   \mathbf{y}_i)}   =   \log{p(\mathbf
{y}_i   ,   \mathbf{q}_i \mid
\mathbf{m}_i)},
\end{equation}
where $g(\mathbf{q}_i) = \log{p(\mathbf{q}_i \mid\mathbf{m}_i)} -
\frac{1}{2}   (T
\log{2\pi}   -   \log{\operatorname{det}(\bolds{\Omega})}   +   \sum
_{t=1}^{T} \mu
^{\star
  2}_{ti}/\sigma^2_t)$. Observe that the linear coefficients
$\mathbf{h}(\mathbf{q}_i) = \bolds{\gamma}_i$ depend on
$\mathbf{q}_i$ through $\mathbf{X}
_{ti}$, while the
concentration matrix $\mathbf{K}(\mathbf{q}_i) = \bolds{\Omega}$
does not. Thus, our model is a
homogeneous CG model [Lauritzen (\citeyear{La1996}), page 160]. Furthermore, since
our genetic model was derived from a set of regression equations
with normal errors, our model is in the homogeneous conditional
Gaussian regression parametric family.

\section{A causal framework for systems genetics}\label{sec3}

This section formalizes our causal inference framework for systems
genetics that combines ideas from randomized experiments with a
purely probabilistic approach to causal inference. We argue that
while causal claims about the relationship between QTLs and
phenotypes are justified by randomization of alleles during meiosis
and the unidirectional influence of genotype on phenotype, causal
claims about the relationships between phenotypes follow from
conditional probability. In a nutshell, by adding QTL nodes to
phenotype networks, we can distinguish between phenotype networks
that would, otherwise, be distribution equivalent.

In order to formalize our approach, we first show that adding causal
QTL nodes can break Markov-equivalence among phenotype networks by
creating new conditional independence relationships among nodes.
Second, we note that two models in the HCGR parametric family are
distribution equivalent if and only if they are Markov equivalent.
The last result together with Theorem \ref{teo1} (see below) provide a simple
graphical criterion to determine whether two DAGs belonging to the
HCGR parametric family are distribution equivalent.

The analogy between the randomization of alleles during meiosis and a
randomized experimental design was first pointed out by Li et al.
(\citeyear{LiTsShStWePaCh2006}). Causality can be unambiguously inferred from a randomized
experiment for two reasons [Dawid (\citeyear{Da2007})]: (1) the treatment to an
experimental unit (genotype) precedes measured outcomes (phenotypes);
and (2) random allocation of treatments to experimental units
guarantees that other common causes are averaged out. The central dogma
of molecular biology [Crick (\citeyear{Cr1958})] suggests that QTL genotypes
generally precede phenotypes. For the types of Eukaryotic data which we
analyze (including clinical traits gene expression and metabolites),
causality must go from QTL to phenotype. Furthermore, random allocation
of QTL genotypes eliminates confounding from other genetic and
environmental effects.

Causal relationships among phenotypes require the additional assumption
of conditional independence. Suppose a QTL, $Q$, and two
phenotypes mapping to $Q$, $Y_1$ and $Y_2$, have true causal
relationship $Q \rightarrow Y_1 \rightarrow Y_2$. That is, $Y_2$ is
independent of $Q$ given $Y_1$. The randomization of genotypes in $Q$
leads to a randomization of $Y_1$, thus averaging out confounding
affects. However, precedence of the `randomized' $Y_1$ before the
`outcome' $Y_2$ cannot in general be determined a priori. Conditional
independence is the key to determine causal order among phenotypes
[Schadt et al. (\citeyear{ScLaYaZhirkt2005}), Li et al. (\citeyear{LiTsShStWePaCh2006}), Chen, Emmert-Streib and Storey (\citeyear{ChEmSt2007}), Liu, de la Fuente and Hoeschele
(\citeyear{LideHo2008}), Chaibub Neto et al. (\citeyear{ChFeAtYa2008})].

In the remainder of this section we present some graphical model
definitions and results needed in the formalization of our causal
graphical models framework. A path is any unbroken, nonintersecting
sequence of edges in a graph, which may go along or against the
direction of the arrows.

\begin{definition}[(d-separation)]
 A path $p$ is said to be d-separated (or
blocked) by a set of nodes $Z$ if and only if
\begin{enumerate}
\item $p$ contains a chain $i \rightarrow m \rightarrow j$ or a fork $i
\leftarrow m \rightarrow j$ such that the middle node \textit{m} is in \textit{Z}, or
\item \textit{p} contains an inverted fork (or collider) $i \rightarrow m
\leftarrow j$ such that the middle node \textit{m} is not in \textit{Z} and such that no
descendant of \textit{m} is in \textit{Z}.
\end{enumerate}
A set \textit{Z} is said to d-separate \textit{X} from \textit{Y} if and only if \textit{Z} blocks every
path from a node in \textit{X} to a node in \textit{Y}. \textit{X} and \textit{Y} are d-connected if
they are not d-separated [Pearl (\citeyear{Pe1988}, \citeyear{Pe2000})].
\end{definition}

Two graphs are Markov equivalent (or faithful indistinguishable) if
they have the same set of d-separation relations [Spirtes, Glymour and Scheines
(\citeyear{SpGlSc2000})]. The skeleton of a causal graph is the undirected graph
obtained by replacing its arrows by undirected edges. A v-structure
is composed by two converging arrows whose tails are not connected
by an arrow.

\begin{theorem}[(Detecting Markov equivalence)]\label{teo1}
Two directed acyclic
graphs (DAGs) are Markov equivalent if and only if they have the
same skeletons and the same set of v-structures [Verma and Pearl
(\citeyear{VwPe1990})].
\end{theorem}

Two models are likelihood equivalent if $f(y \mid M_1)   =   f(y
\mid M_2)$ for any data set $y$, where $f(y \mid M)$ represent the
prior predictive density of the data, $y$, conditional on model $M$
[Heckerman, Geiger and Chickering (\citeyear{HeGeCh1995})]. In this paper we extend the definition of
likelihood equivalence to predictive densities obtained by
plugging in maximum likelihood estimates in the respective sampling
models. A closely related concept, distribution equivalence, states
that two models are distribution equivalent if one is a
reparametrization of the other. While likelihood equivalence is
defined in terms of predictive densities (prior predictive density
or sampling model evaluated on the maximum likelihood estimates),
distribution equivalence is defined in terms of the sampling model
directly. Because of the invariance property of maximum likelihood
estimates, distribution and likelihood equivalence are equivalent
concepts in the frequentist setting. This is also true in the
Bayesian setting with proper priors invariant to model
reparameterizations.

Suppose that for each pair of connected phenotypes in a graph there
exists at least one QTL affecting one but not the other phenotype.
Denote this new graph with QTLs included by the ``extended graph.''
The next result shows that, in this particular situation, we can
distinguish between causal models belonging to a Markov equivalent
class of phenotype networks.

\begin{result}\label{markovlik}
Consider a class of Markov equivalent DAGs $\mathcal{G}$. Let $Y_1$
and $Y_2$ be any two adjacent nodes in the graphs in $\mathcal{G}$.
Assume that for each such pair there exists at least one variable,
$Q$, directly affecting $Y_1$ but not $Y_2$. Let $\mathcal{G}_E$
represent the class of extended graphs. Then the graphs in
$\mathcal{G}_E$ are not Markov equivalent.
\end{result}

As an illustrative example consider the following class of Markov
equivalent models: $\mathcal{G} = \{ Y_1 \rightarrow Y_2 \rightarrow
Y_3   ,   Y_1 \leftarrow Y_2 \leftarrow Y_3   ,   Y_1 \leftarrow
Y_2 \rightarrow Y_3 \}$. These causal models are Markov equivalent
because they have the same set of conditional independence
relations, namely, $Y_1 \perp \!\!\!\!\perp Y_3 \mid Y_2$. In
accordance with Theorem \ref{teo1}, the three models have the same
skeleton, $Y_1 - Y_2 - Y_3$, and the same set of v-structures (no
v-structures). Now consider one QTL, $Q$, affecting $Y_2$ but not
$Y_1$ and $Y_3$. Then $\mathcal{G}_E$ is composed by
\[
\xymatrix@-1pc{
& Q \ar[d] & && & Q \ar[d] & && & Q \ar[d] & \\
Y_1 \ar[r] & Y_2 \ar[r] & Y_3 && Y_1 & Y_2 \ar[l] & Y_3 \ar[l] &&
Y_1 & Y_2 \ar[l] \ar[r] & Y_3. \\
}
\]

Observe that these models still have the same skeleton but different
sets of v-structures: $Y_1 \rightarrow Y_2 \leftarrow Q$, $Q
\rightarrow Y_2 \leftarrow Y_3$ and $\varnothing$, respectively. The
next result guarantees that for the HCGR parametric family Markov
equivalence implies distribution equivalence and vice-versa.

\begin{result}\label{CGmarkovlik}
For the HCGR parametric family, two DAGs are distribution equivalent
if and only if they are Markov equivalent.
\end{result}

It follows from Results \ref{markovlik} and \ref{CGmarkovlik} that by
extending the phenotype network to include QTLs we are able to reduce
the size of and equivalence class of graphs (possibly to a single
network). Furthermore, if we consider Theorem~\ref{teo1} and Result~\ref{CGmarkovlik} together, we
have the following:

\begin{result}\label{HCGRlikeq}
For the HCGR parametric family, two DAGs are distribution equivalent
if and only if they have the same skeletons and same sets of
v-structures.
\end{result}

Result \ref{HCGRlikeq} provides a simple graphical criterion to
determine whether two HCGR models are distribution equivalent. This
allows us to determine distribution equivalence by inspection of
graph structures without the need to go through algebraic
manipulations of joint probability distributions as in Chaibub Neto
et al. (\citeyear{ChFeAtYa2008}).

\section{QTL mapping and phenotype network structure}\label{sec4}

In this section we show that the conditional LOD score can be used as a
formal measure of conditional independence relationships between
phenotypes and QTLs. Even though in this paper we restrict our
attention to HCGR models, conditional LOD profiling is a general
framework for the detection of conditional independencies between
continuous and discrete random variables and does not depend on the
particular parametric family adopted in the modeling. Contrary to
partial correlations, the conditional LOD score does not require the
assumption of multinormality of the data in order to formally test for
independence [recall that only in the Gaussian case, a zero (partial)
correlation implies statistical (conditional) independence], and it can
handle interactive covariates.

The conditional LOD score is defined as
\begin{eqnarray}
\operatorname{LOD}(y   ,   q \mid x) &=& \operatorname{LOD}(y   ,   q   ,   x) -
\operatorname{LOD}(y   ,   x)\nonumber\\[-8pt]\\[-8pt]
& =& \log_{10}\biggl\{ \frac{f(y \mid q   ,   x)}{f(y)}
\biggr\} - \log_{10}\biggl\{ \frac{f(y \mid x)}{f(y)} \biggr\},\nonumber
\end{eqnarray}
where $f(\cdot)$ represents a predictive density (a maximized likelihood
or the prior predictive density in a Bayesian setting). It follows
directly from this definition that
%
\begin{equation}
\quad \operatorname{LOD}(y   ,   q \mid x) = 0 \quad  \Leftrightarrow\quad
 f(y \mid q   ,   x) = f(y \mid x) \quad
\Leftrightarrow\quad  Y \perp \!\!\!\!\perp Q \mid X.
\end{equation}
Therefore, we can use conditional LOD scores as a formal measure of
independence between continuous ($Y$) and discrete ($Q$) random
variables, conditional on any set of variables $X$, that could be
either continuous, discrete or both.

\begin{figure}[b]

\includegraphics{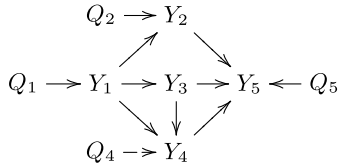}

\caption{Example network with five phenotypes and four QTLs.}
\label{truemodelfig}
\end{figure}

Furthermore, the conditional LOD score can be used to formally test
for conditional independence in the presence of interacting
covariates (denoted by $X \cdot Q$) since
%
\begin{equation}
\qquad \operatorname{LOD}(y   ,   q \mid x   ,   x \cdot q) = \log_{10}\biggl\{
\frac{f(y \mid q   ,   x   ,   x \cdot q)}{f(y)} \biggr\} -
\log_{10}\biggl\{ \frac{f(y \mid x)}{f(y)} \biggr\} = 0
\end{equation}
if and only if $Y \perp \!\!\!\!\perp \{Q   ,   X \cdot Q\} \mid X$.
This is a
very desirable property since, in general, testing for conditional
independence in the presence of interactions is not straightforward.
For example, Andrei and Kendziorski (\citeyear{AnKe2008}) point that in the
presence of interactions, there is no one-to-one correspondence
between zero partial correlations and conditional independencies,
even when we assume normality of the full conditional distributions.

\begin{figure}

\includegraphics{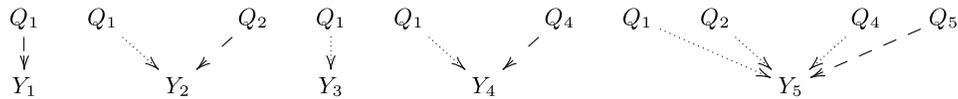}

\caption{Output of a single trait QTL mapping analysis for the
phenotypes in Figure \protect\ref{truemodelfig}. Dashed and pointed arrows
represent direct and indirect QTL/phenotype causal relationships,
respectively.} \label{outsingletraitanalysis}
\end{figure}

Traditional QTL mapping focuses on single trait analysis, where the
network structure among the phenotypes is not taken into
consideration in the analysis. Thus, single-trait analysis may
detect QTLs that directly affect the phenotype under investigation,
as well as QTLs with indirect effects, affecting phenotypes upstream
to the phenotype under study. Consider, for example, the causal
graph in Figure \ref{truemodelfig}. The outputs of single trait
analysis when Figure \ref{truemodelfig} represents the true network
are given in Figure \ref{outsingletraitanalysis}.

\begin{figure}[b]

\includegraphics{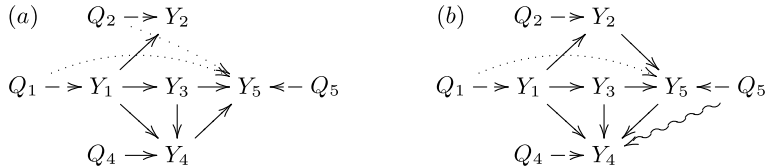}

\caption{QTL mapping tailored to the network structure. \textup{(a)} and \textup{(b)}
display the results of QTL mapping according to slightly altered
network structures from Figure \protect\ref{truemodelfig}. Dashed, pointed and wiggled arrows
represent, respectively, direct, indirect and incorrect
QTL/phenotype causal relationships.} \label{truemodelfigab}
\end{figure}

Now let's consider QTL mapping according to the phenotype network
structure. When the phenotype structure corresponds to the true
causal model, we avoid detecting indirect QTLs by simply performing
mapping analysis of the phenotypes conditional on their parents. For
example, in Figure \ref{truemodelfig}, if we perform a mapping
analysis of $Y_5$ conditional on $Y_2$, $Y_3$ and $Y_4,$ we do not
detect $Q_1$, $Q_2$ and $Q_4$ because $Y_5 \perp \!\!\!\!\perp Q_1
\mid Y_2, Y_3,
Y_4$, $Y_5 \perp \!\!\!\!\perp Q_2 \mid Y_2, Y_3, Y_4$ and $Y_5 \perp
\!\!\!\!\perp Q_4 \mid Y_2,
Y_3, Y_4$. We only detect $Q_5$ since $Y_5 \not\!\perp \!\!\!\!\perp
Q_5 \mid Y_2, Y_3,
Y_4$.

Now consider Figure \ref{truemodelfigab}(a). If we perform a mapping analysis of $Y_5$
conditional on $Y_3$ and $Y_4$, we still detect $Q_1$ and $Q_2$ as QTLs
for $Y_5$, since failing to condition on $Y_2$ leaves the paths $Q_1
\rightarrow Y_1 \rightarrow Y_2 \rightarrow Y_5$ and $Q_2 \rightarrow
Y_2 \rightarrow Y_5$ in Figure \ref{truemodelfig} open. In other words, $Q_1$ and $Q_2$
are d-connected to $Y_5$ conditional on ($Y_3,Y_4$) in the true causal
graph. Mistakenly inferring that a QTL has a direct effect when in
reality it indirectly affects the phenotype is problematic, but not a
serious concern.

On the other hand, if we map an upstream phenotype conditional on
downstream phenotypes, we could infer that downstream QTLs are causal.
This would be a serious problem, as it would dramatically reverse the
causal flow. Consider, for example, Figure \ref{truemodelfigab}(b). If we perform a
mapping analysis of $Y_4$ conditional on $Y_1$, $Y_3$ and~$Y_5$, we
incorrectly detect $Q_5$ as a QTL for $Y_4$ because in the true network
the paths $Y_4 \rightarrow Y_5 \leftarrow Q_5$ and $Y_4 \leftarrow Y_3
\rightarrow Y_5 \leftarrow Q_5$ in Figure \ref{truemodelfig} are open when we condition
on $Y_5$. That is, if we perform mapping analysis of a phenotype
conditional on phenotypes located downstream in the true network, we
induce dependencies between the upstream phenotype and QTLs affecting
downstream phenotypes, and we erroneously conclude that a downstream
QTL affects an upstream phenotype. However, a model with reversed
causal relationships among phenotypes and incorrectly having downstream
QTLs detected as direct QTLs for the upstream node will generally have
a lower marginal likelihood score than the model with the correct
causal order for the phenotypes and correct genetic architecture.
Therefore, in practice, our model selection procedure protects against
this type of mistake.

\section{QTLnet algorithm}\label{sec5}

In this section we propose a statistical framework (\mbox{QTLnet}) for the
joint inference of phenotype network structure and genetic
architecture in systems genetics studies. Work to date in genetical
network reconstruction has treated the problems of QTL inference and
phenotype network reconstruction separately, generally performing
genetic architecture inference first, and then using QTLs to help in
the determination of the phenotype network structure [Chaibub Neto et
al.
(\citeyear{ChFeAtYa2008}), Zhu et al. (\citeyear{ZhZhSmDrirkt2008})]. As indicated in the previous section, such
strategy can incorporate QTLs with indirect effects into the genetic
architecture of phenotypes.

The great challenge in the reconstruction of networks is that the
graph space grows super-exponentially with the number of nodes, so
that exhaustive searches are impractical even for small networks,
and heuristic approaches are needed to efficiently traverse the
graph space. The Metropolis--Hastings algorithm below integrates the
sampling of network structures [Madigan and York (\citeyear{MaYo1995}), Husmeier
(\citeyear{Hu2003})] and QTL mapping.

Let $\mathcal{M}$ represent the structure of a phenotype network
composed of
$T$ nodes. The posterior probability of a specified structure is
given by
%
\begin{equation}
p(\mathcal{M}\mid\mathbf{y}  ,   \mathbf{q}) = \frac{p(\mathbf
{y}\mid\mathbf{q}  ,   \mathcal{M})
p(\mathcal{M})}{\sum_{\mathcal{M}} p(\mathbf{y}\mid\mathbf{q}  ,
  \mathcal{M})   p(\mathcal{M})},
\label{postprob}
\end{equation}
where the marginal likelihood
%
\begin{equation}
p(\mathbf{y}\mid\mathbf{q}  ,   \mathcal{M}) = \int_{\bolds
{\Gamma}} p(\mathbf{y}\mid\mathbf{q}  ,   \bolds{\Gamma}  ,
  \mathcal{M})   p(\bolds{\Gamma}\mid\mathcal{M})  \, d\bolds
{\Gamma}
\end{equation}
is obtained by integrating the product of the prior and likelihood
of the HCGR model with respect to all parameters $\bolds{\Gamma}$ in
the model.
Assuming that the phenotype network is a DAG, the likelihood
function factors according to $\mathcal{M}$ as
%
\begin{equation}
p(\mathbf{y}_i \mid\mathbf{q}_i   ,   \bolds{\Gamma}  ,
\mathcal{M}) = \prod_{t} p(y_{ti}
\mid\mathbf{q}_{ti}   ,   \operatorname{pa}(y_t)),
\end{equation}
where
%
\begin{equation}
p(y_{ti} \mid\mathbf{q}_{ti}   ,   \operatorname{pa}(y_t) ) = N\biggl(
\mu^{\star}_{ti}   +   \sum_{y_k \in \operatorname{pa}(y_t)} \beta_{tk}
y_{ki}   ,   \sigma^2_t \biggr)
\end{equation}
and the problem factors out as a series of linear regression models.
(Note that QTL genotypes $\mathbf{q}_{ti}$ enter the model through
$\mu^{\star}_{ti}$.)

We estimate the posterior probability in (\ref{postprob}) using a
Metropolis--Hastings algorithm detailed in Section 1 of the Supplement.
The M--H proposals, which make single changes (add or drop an edge, or
change causal direction), require remapping of any phenotypes that have
altered sets of parent nodes. The accept/reject calculation involves
estimation of the marginal likelihood conditional on the parent nodes
and newly mapped QTL(s).

Because the graph space grows rapidly with the number of phenotype
nodes, the network structure with the highest posterior probability
may still have a very low probability. Therefore, instead of
selecting the network structure with the highest posterior
probability, we perform Bayesian model averaging [Hoeting et al.
(\citeyear{HoMaRaVo1999})] for the causal relationships between phenotypes
and infer an
averaged network. Explicitly, let $\Delta_{uv}$ represent a causal
relationship between phenotypes $u$ and~$v$, that is, $\Delta_{uv} =
\{ Y_u \rightarrow Y_v   ,   Y_u \leftarrow Y_v   ,   Y_u
\not\rightarrow Y_v   \mbox{ and }   Y_u \not\leftarrow Y_v \}$.
Then
\begin{eqnarray}\label{bayesmodelaveraging}
p(\Delta_{uv} \mid\mathbf{y})  & =&   \sum_{k} p(\Delta_{uv} \mid
\mathcal{M}_k
,   \mathbf{y}  ,   \mathbf{q})   p(\mathcal{M}_k \mid\mathbf
{y}  ,   \mathbf{q})\nonumber\\[-8pt]\\[-8pt]
& =&   \sum_k \mathbh{1}\{ \Delta_{uv} \in\mathcal
{M}_k \}
p(\mathcal{M}_k \mid\mathbf{y}  ,   \mathbf{q}) .\nonumber
\end{eqnarray}
The averaged network is constructed by putting together all causal
relationships with maximum posterior probability or with posterior
probability above a predetermined threshold.

\section{Simulations}\label{sec6}

In this section we evaluate the performance of the \mbox{QTLnet} approach
in simulation studies of a causal network with five phenotypes and
four causal QTLs. We consider situations with strong or weak causal
signals, leading respectively to high or low phenotype correlations.
We show that important features of the causal network can be
recovered. Further, this simulation illustrates how an alleged
hot spot could be explained by sorting out direct and indirect
effects of the~QTLs.

We generated 1000 data sets according to Figure \ref{truemodelfig}.
Each simulated cross object [Broman et al. (\citeyear{BrWuSeCh2003})] had 5 phenotypes
simulated for an F2 population with 500 individuals. The genome had
5 chromosomes of length 100 cM with 10 equally spaced markers per
chromosome. We simulated one QTL per phenotype, except for phenotype
$Y_3$ with no QTLs. The QTLs $Q_t$, $t=1,2,4,5,$ were unlinked and
placed at the middle marker on chromosome $t$.

Each simulated cross object had different sampled parameter value
combinations for each realization. In the strong signal simulation,
we sampled the additive and dominance effects according to
$U[0.5,1]$ and $U[0,0.5]$, respectively. The partial regression
coefficients for the phenotypes were sampled according to
$\beta_{uv} \sim0.5   U[-1.5,-0.5] + 0.5   U[0.5,1.5]$. In the
weak signal simulation, we generated data sets with additive and
dominance effects from $U[0,0.5]$ and $U[0,0.25]$, respectively, and
partial regression coefficients sampled with $\beta_{uv} \sim
U[-0.5,0.5]$. The residual phenotypic variance was fixed at 1 in
both settings.

We first show the accuracy of the mapping analysis in our simulated
data sets. We used interval mapping with a LOD score threshold of 5
to detect significant~QTLs. Table \ref{condmapnetstr} shows the
results of both unconditional and QTL mapping according to the
phenotype network in Figure \ref{truemodelfig}. In the strong signal
setting, the unconditional mapping often detected indirect QTLs, but
the mapping of phenotypes conditional on their parent nodes
increased detection of the true genetic architectures. In the weak
signal simulation, the unconditional mapping did not detect indirect
QTLs in most cases, but we still observe improvement in detection of
the correct genetic architecture when we condition on the parents.

The expected architecture contains the d-connected QTLs when
conditioning (or not) on other phenotypes as indicated in the first
column of Table \ref{condmapnetstr}. For instance, $Q_1$ and $Q_2$ are
d-connected to $Y_2$, but only $Q_2$ is d-connected to $Y_2$ when
properly conditioning on $Y_1$. Supplementary Tables S1, S2, S3, S4 and
S5 show the simulation results for all possible conditional mapping
combinations.

\begin{table}
\tabcolsep=0pt
\caption{Frequencies of QTL detection for both unconditional (top
half) and conditional (bottom half) QTL mapping according with the
true phenotype network structure in Figure \protect\ref{truemodelfig}.
Results, for each simulation, based on 1000 simulated data sets
described in the text. The expected architecture~is~the~set of
d-connected QTLs for the phenotype conditioning with
respect~to~the~network in Figure \protect\ref{truemodelfig}}
\label{condmapnetstr}
\begin{tabular*}{\textwidth}{@{\extracolsep{\fill}}lccccccccc@{}}
\hline
& \multicolumn{4}{c}{\textbf{Strong signal}} & \multicolumn{4}{c}{\textbf{Weak signal}} & \multirow{2}{50pt}[-7pt]{\centering \textbf{Expected architecture}}\\[-6pt]
& \multicolumn{4}{c}{\hrulefill} & \multicolumn{4}{c}{\hrulefill} & \\
\textbf{Phenotypes} & $\bolds{Q_1}$ & $\bolds{Q_2}$ & $\bolds{Q_4}$ & $\bolds{Q_5}$ & $\bolds{Q_1}$ & $\bolds{Q_2}$ & $\bolds{Q_4}$ &$\bolds{Q_5}$ &  \\
\hline
$Y_{1}$ & 0.997 & 0.000 & 0.000 & 0.000 & 0.431 & 0.000 & 0.000 & 0.000& $\{ Q_{1} \}$ \\
$Y_{2}$ & 0.884 & 0.930 & 0.000 & 0.000 & 0.001 & 0.384 & 0.000 & 0.000& $\{ Q_{1}, Q_2 \}$ \\
$Y_{3}$ & 0.941 & 0.000 & 0.000 & 0.000 & 0.003 & 0.000 & 0.000 & 0.000& $\{ Q_{1} \}$ \\
$Y_{4}$ & 0.603 & 0.000 & 0.690 & 0.000 & 0.003 & 0.000 & 0.370 & 0.000& $\{ Q_{1}, Q_4 \}$ \\
$Y_{5}$ & 0.637 & 0.321 & 0.321 & 0.340 & 0.000 & 0.000 & 0.001 & 0.336& $\{ Q_{1}, Q_2, Q_4, Q_5 \}$ \\
[6pt]
$Y_{2} \mid Y_{1}$ & 0.001 & 0.999 & 0.000 & 0.000 & 0.000 & 0.424 &0.000 & 0.000 & $\{ Q_{2} \}$ \\
$Y_{3} \mid Y_{1}$ & 0.000 & 0.000 & 0.000 & 0.000 & 0.000 & 0.000 &0.000 & 0.000 & $\varnothing$ \\
$Y_{4} \mid Y_{1}, Y_{3}$ & 0.000 & 0.000 & 0.999 & 0.000 & 0.000 &0.000 & 0.422 & 0.000 & $\{ Q_{4} \}$ \\
$Y_{5} \mid Y_{2}, Y_{3}, Y_{4}$ & 0.000 & 0.000 & 0.000 & 0.999 &0.000 & 0.000 & 0.000 & 0.415 & $\{ Q_{5} \}$ \\
\hline
\end{tabular*}
\end{table}

For each simulated data set we applied the \mbox{QTLnet} algorithm using
simple interval mapping for QTL detection. The ratio of marginal
likelihoods in the Metropolis--Hastings algorithm was computed using
the BIC asymptotic approximation to the Bayes factor (equation 1.1 on
the Supplement). We adopted
uniform priors over network structures. We ran each Markov chain for
30,000 iterations, sampled a structure every 10 iterations, and
discarded the first 300 (burnin) network structures producing
posterior samples of size 2700. Posterior probabilities for each
causal relationship were computed via Bayesian model averaging.

Table \ref{posteriorrank} shows the frequency, out of the 1000
simulations, the true model was the most probable, second most
probable, etc. The results show that in the strong signal setting
the true model got the highest posterior probability in most of the
simulations. In the weak signal setting the range of rankings was
very widespread.\looseness=1

\begin{table}
\tabcolsep=0pt
\caption{Frequencies that the posterior probability of the true
model was the highest, second highest, etc. Results based on 1000
simulated data sets described in the text} \label{posteriorrank}
\begin{tabular*}{\textwidth}{@{\extracolsep{\fill}}lccccccccccccc@{}}
\hline
& \textbf{1st} & \textbf{2nd} & \textbf{3rd} & \textbf{4th} & \textbf{5th} & \textbf{6th} & \textbf{7th} & \textbf{8th} & \textbf{9th} & \textbf{10th}
& \textbf{11th} &\textbf{12th} & $\bolds{\geq}$\textbf{13th} \\
\hline
Strong & 842 & 100 & 21 & 11 & \phantom{0}3 & \phantom{0}4 & \phantom{0}3 & \phantom{0}2 & 1 & \phantom{0}1 & \phantom{0}3 & 1 & \phantom{00}8 \\
Weak & \phantom{0}21 & \phantom{0}33 & 18 & 19 & 19 & 16 & 17 & 15 & 8 & 12 & 13 & 5 & 804 \\
\hline
\end{tabular*}
\end{table}

Table \ref{mavfreq} shows the proportion of times that each possible
causal direction ($Y_u \rightarrow Y_v$,  $Y_u \leftarrow Y_v$) or
no causal relation ($\{Y_u \not\rightarrow Y_v   ,   Y_u
\not\leftarrow Y_v\}$) had the highest posterior probability for all
pairs of phenotypes. The results show that in the strong signal
simulations, the correct causal relationships were recovered with
high probability. The results are weaker but in the correct
direction in the weak signal setting.\looseness=1

\begin{table}[b]
\caption{Frequencies that each possible causal relationship had the
highest posterior probability (computed via Bayesian model
averaging). Results based on 1000 simulated data sets described in
the text} \label{mavfreq}
\begin{tabular*}{\textwidth}{@{\extracolsep{\fill}}lcccccc@{}}
\hline
& \multicolumn{3}{c}{\textbf{Strong signal}} & \multicolumn{3}{c@{}}{\textbf{Weak signal}} \\[-6pt]
& \multicolumn{3}{c}{\hrulefill} & \multicolumn{3}{c@{}}{\hrulefill} \\
\textbf{Phenotypes} & $\bolds{\rightarrow}$ & $\bolds{\leftarrow}$ & $\bolds{\not\rightarrow, \not\leftarrow}$ & $\bolds{\rightarrow}$ & $\bolds{\leftarrow}$ & $\bolds{\not\rightarrow,
\not\leftarrow}$ \\
\hline
(1, 2) & 0.996 & 0.002 & 0.002 & 0.594 & 0.177 & 0.229 \\
(1, 3) & 0.990 & 0.002 & 0.008 & 0.471 & 0.263 & 0.266 \\
(1, 4) & 0.990 & 0.001 & 0.009 & 0.541 & 0.196 & 0.263 \\
(1, 5) & 0.054 & 0.002 & 0.944 & 0.028 & 0.005 & 0.967 \\
(2, 3) & 0.016 & 0.022 & 0.962 & 0.018 & 0.017 & 0.965 \\
(2, 4) & 0.037 & 0.012 & 0.951 & 0.018 & 0.015 & 0.967 \\
(2, 5) & 0.997 & 0.003 & 0.000 & 0.712 & 0.075 & 0.213 \\
(3, 4) & 0.967 & 0.031 & 0.002 & 0.482 & 0.253 & 0.265 \\
(3, 5) & 0.997 & 0.003 & 0.000 & 0.653 & 0.116 & 0.231 \\
(4, 5) & 0.996 & 0.004 & 0.000 & 0.670 & 0.115 & 0.215 \\
\hline
\end{tabular*}
\end{table}

Interestingly, single trait analysis with strong signal showed that
$Y_5$ mapped to~$Q_1$ more frequently than to $Q_5$ (Table
\ref{condmapnetstr}). This result can be understood using a path
analysis [Wright (\citeyear{Wr1934})] argument. In path analysis, we decompose the
correlation between two variables among all paths connecting the two
variables in a graph. Let $\mathcal{D}_{uv}$ represent the set of
all direct and indirect directed paths connecting $u$ and $v$ (a
directed path is a path with all arrows pointing in the same
direction). Then the correlation between these nodes can be
decomposed as
\begin{eqnarray}\label{pathcoeffs}
\operatorname{cor}(y_u,y_v) &=& \sum_{P   \in  \mathcal{D}_{uv}} \phi_{p_2 u}
\phi_{p_3 p_2} \cdots\phi_{v p_{m-1}}\nonumber\\[-8pt]\\[-8pt]
&=&   \biggl\{\frac{\operatorname{var}(y_u)}{\operatorname{var}(y_v)}\biggr\}^{1/2}
\sum_{P   \in
\mathcal{D}_{uv}} \beta_{p_2 u}   \beta_{p_3 p_2} \cdots\beta_{v
p_{m-1}},\nonumber
\end{eqnarray}
where $\phi_{ij} = \beta_{ij}
\{{\operatorname{var}(y_j)}/{\operatorname{var}(y_i)}\}^{1/2}$ is a standardized
path coefficient. Assuming intra-locus additivity and encoding the
genotypes as 0, 1 and 2 (for the sake of easy computation), we have
from equation (\ref{pathcoeffs}) that
\begin{eqnarray*}
\operatorname{cor}(y_5,q_1) &=& \beta_{1,q_1} ( \beta_{52}   \beta_{21} + \beta_{53}
  \beta_{31} + \beta_{54}   \beta_{41} + \beta_{54}   \beta_{43}
  \beta_{31} )
  \biggl\{\frac{\operatorname{var}(q_1)}{\operatorname{var}(y_5)}\biggr\}^{1/2},\\
\operatorname{cor}(y_5,q_5) &=& \beta_{1,q_5}
\biggl\{\frac{\operatorname{var}(q_5)}{\operatorname{var}(y_5)}\biggr\}^{1/2}.
\end{eqnarray*}
We therefore see that if the partial regression coefficients between
phenotypes are high, and the QTL effects $\beta_{1,q_1}$ and
$\beta_{5,q_5}$ and QTL variances are close (as in the strong signal
simulation), then $\operatorname{cor}(y_5,q_1)$ will be higher than $\operatorname{cor}(y_5,q_5)$
and $Y_5$ will map to $Q_1$ with stronger signal than to $Q_5$.

This result suggests a possible scenario for the appearance of eQTL
hot spots when phenotypes are highly correlated. A set of correlated
phenotypes may be better modeled in a causal network with one
upstream phenotype that in turn has a causal QTL. Ignoring the
phenotype network can result in an apparent hot spot for the
correlated phenotypes. Here, all phenotypes detect QTL $Q_1$ with
high probability when mapped unconditionally (Table
\ref{condmapnetstr}). In the weak signal setting, the phenotypes map
mostly to their respective QTLs and do not show evidence for a
hot spot. No hot spot was found in additional simulations having
strong QTL/phenotype relationships and weak phenotype/phenotype
relations (results not shown). Thus, our \mbox{QTLnet} approach can
effectively explain a hot spot found with unconditional mapping when
phenotypes show strong causal structure.

\section{Network inference for a liver hot spot}\label{sec7}

In this section we illustrate the application of \mbox{QTLnet} to a subset
of gene expression data derived from a F2 intercross between inbred
lines $C3H/HeJ$ and $C57BL/6J$ [Ghazalpour et al. (\citeyear{GhDoZhWairkt2006}), Wang et al.
(\citeyear{WaYeScWaDrLu})]. The data set is composed of genotype data on 1,065 markers
and liver expression data on the 3421 available transcripts from 135
female mice. Interval mapping indicates that 14 transcripts map to
the same region on chromosome 2 with a LOD score higher than 5.3
(permutation $p$-value $<0.001$). Only one transcript, $\mathit{Pscdbp}$, is
located on chromosome 2 near the hot spot locus. The~14 transcripts
show a strong correlation structure, and the correlation structure
adjusting for the peak marker on chromosome 2, $rs3707138$, is still
strong (see Supplementary Table S6). This co-mapping
suggest all transcripts are under the regulation of a common factor.
Causal relationships among phenotypes could explain the strong
correlation structure that we observe, although other possibilities
are environmental factors or a latent factor that is not included.

We applied the \mbox{QTLnet} algorithm on the 129 mice that had no missing
transcript data using Haley--Knott (\citeyear{HaKn1992}) regression (and assuming
genotyping error rate of~0.01) for the detection of QTLs conditional
on the network structure, and we used the BIC approximation to estimate
the marginal likelihood
ratio in the Metropolis--Hastings algorithm (equation
1.1 on the Supplement). We adopted uniform priors over network
structures. We ran a Markov chain for 1,000,000 iterations and
sampled a network structure every 100 iterations, discarding the
first 1000 and basing inference on a posterior sample with 9000
network structures. Diagnostic plots and measures (see Section 3 on the
Supplement) support the convergence of the Markov chain.

We performed Bayesian model averaging and, for each of 91 possible
pairs ($Y_u,Y_v$), we obtained the posterior probabilities of $Y_u
\rightarrow Y_v$, $Y_u \leftarrow Y_v$ and of no direct causal
connection. The results are shown in Supplementary Table~S7. Figure~\ref{averagednetworkislet10} shows a
model-averaged network.

\begin{figure}[b]

\includegraphics{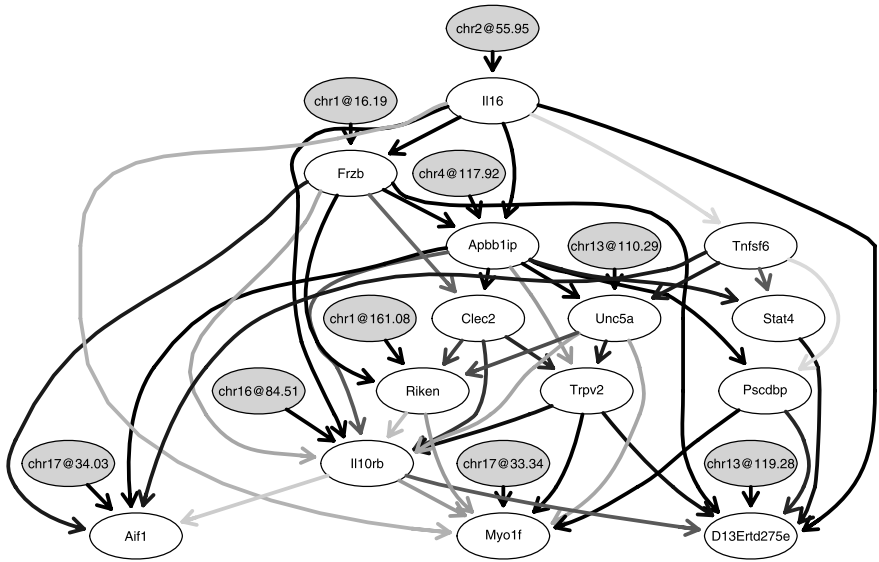}

\caption{Model-averaged posterior network. Arrow darkness is
proportional to the posterior probability of the causal relation
computed via Bayesian model averaging. For each pair of phenotypes, the
figure displays the causal relationship (presence or absence of an
arrow) with highest posterior probability. Light grey nodes represent
QTLs and show their chromosome number and position in centimorgans.
Riken represents the riken gene $6530401C20Rik$.}
\label{averagednetworkislet10}
\end{figure}

This network suggests a key role of $Il16$ in the regulation of the
other transcripts in the network. $Il16$ is upstream to all other
transcripts, and is the only one directly mapping to the locus of
chromosome 2. We would have expected the $\mathit{cis}$ transcript, $\mathit{Pscdbp}$,
to be the upstream phenotype in this network. However, the data
suggests $\mathit{Pscdbp}$ is causal to only two other transcripts and that
some other genetic factor on chromosome 2 may be driving this
pathway. This estimated \mbox{QTLnet} causal network provides new
hypotheses that could be tested in future mouse experiments.

\section{Discussion}\label{sec8}

We have developed a statistical framework for causal inference in
systems genetics. Causal relationships between QTLs and phenotypes
are justified by the randomization of alleles during meiosis
together with the unidirectional influence of genotypes on
phenotypes. Causal relationships between phenotypes follows from
breakage of distribution equivalence due to QTL nodes augmenting the
phenotype network. We have proposed a novel approach to jointly
infer genetic architecture and causal phenotype network structure
using HCGR models. We argue in this paper that failing to properly
account for phenotype network structure for mapping analysis can
yield QTLs with indirect effects in the genetic architecture, which
can decrease the power to detect the correct causal relationships
between phenotypes.

Current literature in systems genetics [Chaibub Neto et al. (\citeyear{ChFeAtYa2008}),
Zhu et al. (\citeyear{ZhZhSmDrirkt2008})] has considered the problems of genetic architecture
and phenotype network structure inference separately. Chaibub Neto
et al. (\citeyear{ChFeAtYa2008}) used the \mbox{PC-algorithm} [Spirtes, Glymour and Scheines (\citeyear{SpGlSc2000})] to first
infer the skeleton of the phenotype network and then use QTLs to
determine the directions of the edges in the phenotype network. Zhu
et al. (\citeyear{ZhZhSmDrirkt2008}) reconstructed networks from a consensus of Bayesian
phenotype networks with a prior distribution based on causal tests
of Schadt et al. (\citeyear{ScLaYaZhirkt2005}). Their prior was computed with QTLs
determined by single trait analysis.

Liu, de la Fuente and Hoeschele (\citeyear{LideHo2008}) presented an approach based in structural equation
models (and applicable to species where sequence information is
available) that partially accounts for the phenotype network
structure when selecting the QTLs to be incorporated in the network.
They perform eQTL mapping using $\mathit{cis}$, $\mathit{cis}$-$\mathit{trans}$ and
$\mathit{trans}$-regulation [Doss et al. (\citeyear{DoScDrLu2005}), Kulp and Jagalur (\citeyear{KuJa2006})] and
then use local structural models to identify regulator-target pairs
for each eQTL. The identified relationships are then used to
construct an encompassing directed network (EDN) with nodes composed
by transcripts and eQTLs and arrows from (1) eQTls to
$\mathit{cis}$-regulated target transcripts; (2) $\mathit{cis}$-regulated transcripts
to $\mathit{cis}$-$\mathit{trans}$-regulated target transcripts; and (3)
$\mathit{trans}$-regulator transcripts to target transcripts, and from
$\mathit{trans}$-eQTL to target transcripts. The EDN defines a network search
space for network inference with model selection based on penalized
likelihood scores and an adaptation of Occam's window [Madigan and
Raftery (\citeyear{MaRa1994})]. Their local structural models, which fit at most two
candidate regulators per target transcript, can include indirect
eQTLs in the genetic architecture of target transcripts when there
are multiple paths connecting a $\mathit{cis}$-regulator to a
$\mathit{cis}$-$\mathit{trans}$-target transcript. In other words, some transcripts
identified as $\mathit{cis}$-regulated targets may actually be $\mathit{cis}$-$\mathit{trans}$.

Winrow et al. (\citeyear{WiWiKaMiirkt2009}) rely on a (nonhomogeneous) conditional Gaussian
model and employ the standard Metropolis--Hastings (M--H) algorithm
(add, remove or delete edges) to search over the space of DAG
structures. It differs from our approach in how QTL detection is
coupled with the M--H proposal. \mbox{QTLnet} constructs M--H proposals for
edges connecting phenotypes and detects QTLs conditional on the network
structure. On the other hand, Winrow et al. first detect QTLs and then
construct M--H proposals for both phenotypes and QTL nodes. In Section 4
of the Supplement we present a simulation study comparing our approach
to Winrow's strategy. \mbox{QTLnet} inferred the proper genetic architecture
and phenotype network structure at a higher rate than Winrow's approach
with high signal-to-noise ratio. In the weak signal setting, \mbox{QTLnet}
picks up direct causal relationship at a higher rate, but Winrow's is
better when there is no direct causal link; on average, they were
comparable. Although Winrow's strategy also avoids the detection of
indirect QTLs, it is prone to miss direct QTLs when the causal effects
among phenotypes are strong, and can perform poorly depending on the
structure of the phenotype network (see Supplement for details).

Current interest in the eQTL literature centers on understanding the
relationships among expression traits that co-map to a genomic
region. It is often suggested that these eQTL ``hot spots'' result
from a master regulator affecting the expression levels of many
other genes [see Breitling et al. (\citeyear{BrLiTeirkt2008})]. A path analysis argument
suggests that if the correlation structure between the phenotypes is
strong because of a strong causal signal, a~well-defined hot spot
pattern will likely appear when we perform single trait analysis.
Our simulations and real data example suggest that this is the
situation where the \mbox{QTLnet} algorithm is expected to be most
fruitful.

The \mbox{QTLnet} approach is based on a Metropolis--Hastings algorithm that
at each step proposes a slightly modified phenotype network and fits
the genetic architecture conditional on this proposed network.
Conditioning on the phenotype network structure should generally
lead to a better inferred genetic architecture. Likewise, a~better
inferred genetic architecture should lead to a better inferred
phenotype structure (models with better inferred genetic
architectures should have higher marginal likelihood scores. A~poorly inferred genetic architecture may compromise the marginal
likelihood of a network with phenotype structure close to the true
network).

Because the proposal mechanism of the Metropolis--Hastings algorithm
is based in small modification of the last accepted network
(addition, deletion or reversion of a single edge), the mixing of
the Markov chain is generally slow and it is necessary to run long
chains and use big thinning windows in order to achieve good mixing.
This is a bottleneck to the scalability of this approach. We
therefore plan to investigate more efficient versions of the
Metropolis--Hastings algorithm for network structure inference. In
particular, a new and more extensive edge reversal move proposed by
Grzegorczyk and Husmeier (\citeyear{GrHu2008}) and an approach based in a Markov
blanket decomposition of the network [Riggelsen (\citeyear{Ri2005})].

Our method is currently implemented with R code. The analysis of the
real data example took over 20 hours in a 64 bit Intel(R) Core(TM) 2
Quad 2.66 GHz machine. We can handle up to 20 phenotype nodes, at this
point. The run time for the code, written in R, should be substantially
improved as we optimize code, converting key functions to C (under
development). Nonetheless, because the number of DAGs increases
super-exponentially with the number of phenotype nodes, scaling up the
proposed approach to large networks will likely be a very challenging
task. Current R code is available from the authors upon request.

One of the most attractive features of a Bayesian framework is its
ability to formally incorporate prior information in the analysis.
Given the complexity of biological processes and the many
limitations associated with the partial pictures provided by any of
the ``omic'' data sets now available, incorporation of external
information is highly desirable. We are currently working in the
development of priors for network structures.

The \mbox{QTLnet} approach can be seen as a method to infer causal Bayesian
networks composed of phenotype and QTL nodes. Standard Bayesian
networks provide a compact representation of the conditional
dependency and independencies associated with a joint probability
distribution. The main criticism of a causal interpretation of such
networks is that different structures may be likelihood equivalent
while representing totally different causal process. In other words,
we can only infer a class of likelihood equivalent networks. We have
formally shown how to break likelihood equivalence by incorporating
causal QTLs.

We have focussed on experimental crosses with inbred founders, as the
recombination model and genetic architecture are relatively
straightforward. However, this approach might be extended to outbred
populations with some additional work. The genotypic effects are
random, and the problem needs to be recast in terms of variance
components.

In this paper we only consider directed acyclic graphs. We point out,
however, that the HCGR parametric family accommodates cyclic networks.
The \mbox{QTLnet} approach can only infer causality among phenotypes that are
consistent with the assumption of no latent variables and no
measurement error. However, these complications can impact network
reconstruction. We are currently investigating extensions of the
proposed framework along these lines.

\section*{Acknowledgments}
We would like to thank Alina Andrei, the
associate editor and two anonymous referees for their helpful comments
and suggestions.

\begin{supplement}
\stitle{Supplement to ``Causal graphical models in systems genetics:
 A unified framework for joint inference of causal network and genetic architecture for correlated phenotypes''}
\slink[doi]{10.1214/09-AOAS288SUPP}
\slink[url]{http://lib.stat.cmu.edu/aoas/288/supplement.pdf}
\sdatatype{.pdf}
\sdescription{The Supplement article presents: (1) the
Metropolis--Hastings algorithm for \mbox{QTLnet}; (2) supplementary tables for
the simulations and real data example; (3) convergence diagnostics for
the real data example; (4) comparison with Winrow et al. (\citeyear{WiWiKaMiirkt2009});
and~(5)~the proofs of Results \ref{concmatrixformula}, \ref{markovlik}, \ref{CGmarkovlik} and \ref{HCGRlikeq}.}
\end{supplement}

%

\printaddresses


\begin{thebibliography}{99}
\bibitem[\protect\citeauthoryear{}{2009}]{AnKe2008}
\textsc{Andrei, A.} and \textsc{Kendziorski, C.} (2009). An efficient
method for identifying statistical interactors in graphical models.
 \textit{Biostatistics} \textbf{10} 706--718.

\bibitem[\protect\citeauthoryear{}{2008}]{AtFuLu2008}
\textsc{Aten, J. E., Fuller, T. F., Lusis, A. J.} and \textsc
{Horvath, S.}
(2008). Using genetic markers to orient the edges in quantitative trait
networks: The NEO software. \textit{BMC Sys. Biol.} \textbf{2} 34.

\bibitem[\protect\citeauthoryear{}{2008}]{BaYaYi2008}
\textsc{Banerjee, S., Yandell, B. S.} and \textsc{Yi, N.} (2008).
Bayesian QTL mapping for multiple traits. \textit{Genetics} \textbf
{179} 2275--2289.

\bibitem[\protect\citeauthoryear{}{2008}]{BrLiTeirkt2008}
\textsc{Breitling, R., Li, Y., Tesson, B. M., Fu, J., Wu, C.,
Wiltshire, T., Gerrits, A., Bystrykh, L. V., de Haan, G., Su, A. I.} and
\textsc{Jansen, R. C.} (2008). Genetical genomics: Spotlight on QTL
hotspots. \textit{PLoS Genet.} \textbf{4} e1000232.

\bibitem[\protect\citeauthoryear{}{2003}]{BrWuSeCh2003}
\textsc{Broman, K., Wu, H., Sen, S.} and \textsc{Churchill, G. A.}
(2003). R/qtl: QTL mapping in experimental crosses. \textit
{Bioinformatics} \textbf{19} 889--890.

\bibitem[\protect\citeauthoryear{}{2008}]{ChFeAtYa2008}
\textsc{Chaibub Neto, E., Ferrara, C., Attie, A. D.} and \textsc
{Yandell, B. S.} (2008). Inferring causal phenotype networks from
segregating populations. \textit{Genetics} \textbf{179} 1089--1100.

\bibitem[\protect\citeauthoryear{}{2009}]{ChKeAtYa2009}
\textsc{Chaibub Neto, E., Keller, M. P., Attie, A. D.} and \textsc
{Yandell, B. S.} (2009). Supplement to
``Causal graphical models in systems genetics: A unified framework for
joint inference of causal network and
genetic architecture for correlated phenotypes.''
DOI: \href{http://dx.doi.org/10.1214/09-AOAS288SUPP}{10.1214/09-AOAS288SUPP}.

\bibitem[\protect\citeauthoryear{}{2007}]{ChEmSt2007}
\textsc{Chen, L. S., Emmert-Streib, F.} and \textsc{Storey, J. D.}
(2007). Harnessing naturally randomized transcription to infer
regulatory relationships among genes. \textit{Genome Biology} \textbf
{8} R219.

\bibitem[\protect\citeauthoryear{}{1958}]{Cr1958}
\textsc{Crick, F. H. C.} (1958). On Protein Synthesis. \textit{Symp.
Soc. Exp. Biol.} \textbf{XII} 139--163.

\bibitem[\protect\citeauthoryear{}{2007}]{Da2007}
\textsc{Dawid, P.} (2007). Fundamentals of statistical causality.
Research Report 279, Dept. Statistical Science, Univ. College London.

\bibitem[\protect\citeauthoryear{}{2005}]{DoScDrLu2005}
\textsc{Doss, S., Schadt, E. E., Drake, T. A., Lusis, A. J.} (2005).
Cis-acting expression
quantitative trait loci in mice. \textit{Genome Research} \textbf{15}
681--691.

\bibitem[\protect\citeauthoryear{}{2006}]{GhDoZhWairkt2006}
\textsc{Ghazalpour, A., Doss, S., Zhang, B., Wang, S., Plaisier, C.,
Castellanos, R., Brozell, A., Schadt, E. E., Drake, T. A., Lusis, A.
J.} and \textsc{Horvath, S.} (2006). Integrating genetic and network
analysis to characterize genes related to mouse weight. \textit{PLoS
Genetics} \textbf{2} e130.

\bibitem[\protect\citeauthoryear{}{2008}]{GrHu2008}
\textsc{Grzegorczyk, M.} and \textsc{Husmeier, D.} (2008). Improving
the structure MCMC sampler for Bayesian networks by introducing a new
edge reversal move. \textit{Machine Learning} \textbf{71} 265--305.

\bibitem[\protect\citeauthoryear{}{1992}]{HaKn1992}
\textsc{Haley, C.} and \textsc{Knott, S.} (1992). A simple regression
method for mapping quantitative trait loci in line crosses using
flanking markers. \textit{Heredity} \textbf{69} 315--324.

\bibitem[\protect\citeauthoryear{}{1995}]{HeGeCh1995}
\textsc{Heckerman, D., Geiger, D.} and \textsc{Chickering, D.}
(1995). Learning Bayesian networks: The~combination of knowledge and
statistical data. \textit{Machine Learning} \textbf{20} 197--243.

\bibitem[\protect\citeauthoryear{}{1999}]{HoMaRaVo1999}
\textsc{Hoeting, J. A., Madigan, D., Raftery, A. E.} and \textsc
{Volinsky, C. T.} (1999). Bayesian model averaging: A tutorial (with discussion and rejoinder by
authors).
\textit{Statist. Sci.} \textbf{14} 382--417.
\MR{1765176}

\bibitem[\protect\citeauthoryear{}{2003}]{Hu2003}
\textsc{Husmeier, D.} (2003). Sensitivity and specificity of inferring
genetic regulatory interactions from microarray experiments with
dynamic Bayesian networks. \textit{Bioinformatics} \textbf{19} 2271--2282.

\bibitem[\protect\citeauthoryear{}{2006}]{KuJa2006}
\textsc{Kulp, D. C.} and \textsc{Jagalur, M.} (2006). Causal
inference of regulator-target pairs by gene mapping of expression
phenotypes. \textit{BMC Genomics} \textbf{7} 125.

\bibitem[\protect\citeauthoryear{}{1996}]{La1996}
\textsc{Lauritzen, S.} (1996). \textit{Graphical Models}. \textit{Oxford Statistical Science Series} \textbf{17}. Oxford
Univ. Press, New York.
\MR{1419991}

\bibitem[\protect\citeauthoryear{}{2006}]{LiTsShStWePaCh2006}
\textsc{Li, R., Tsaih, S. W., Shockley, K., Stylianou, I. M.,
Wergedal, J., Paigen, B.} and \textsc{Churchill, G. A.} (2006).
Structural model analysis of multiple quantitative traits. \textit
{PLoS Genetics} \textbf{2} e114.

\bibitem[\protect\citeauthoryear{}{2008}]{LideHo2008}
\textsc{Liu, B., de la Fuente, A.} and \textsc{Hoeschele, I.} (2008).
Gene network inference via structural equation modeling in genetical
genomics experiments. \textit{Genetics} \textbf{178} 1763--1776.

\bibitem[\protect\citeauthoryear{}{1994}]{MaRa1994}
\textsc{Madigan, D.} and \textsc{Raftery, J.} (1994). Model selection
and accounting for model uncertainty
in graphical models using Occam's window. \textit{J. Amer. Statist.
Assoc.} \textbf{89} 1535--1546.

\bibitem[\protect\citeauthoryear{}{1995}]{MaYo1995}
\textsc{Madigan, D.} and \textsc{York, J.} (1995). Bayesian graphical
models for discrete data. \textit{Int. Stat. Rev.}
\textbf{63} 215--232.

\bibitem[\protect\citeauthoryear{}{1988}]{Pe1988}
\textsc{Pearl, J.} (1988). \textit{Probabilistic Reasoning in
Intelligent Systems: Networks of Plausible Inference}. Kaufmann, San Mateo, CA.
\MR{0965765}

\bibitem[\protect\citeauthoryear{}{2000}]{Pe2000}
\textsc{Pearl, J.} (2000). \textit{Causality: Models, Reasoning and
Inference}. Cambridge Univ. Press, New York.
\MR{1744773}

\bibitem[\protect\citeauthoryear{}{2005}]{Ri2005}
\textsc{Riggelsen, C.} (2005). MCMC learning of Bayesian network
models by Markov blanket decomposition. In \textit{Lecture Notes in
Computer Science} 329--340. Springer, Berlin.

\bibitem[\protect\citeauthoryear{}{2005}]{ScLaYaZhirkt2005}
\textsc{Schadt, E. E., Lamb, J., Yang, X., Zhu, J., Edwards, S.,
Guhathakurta, D., Sieberts, S. K.,
Monks, S., Reitman, M., Zhang, C., Lum, P. Y., Leonardson, A.,
Thieringer, R., Metzger, J. M., Yang, L.,
Castle, J., Zhu, H., Kash, S. F., Drake, T. A., Sachs, A.} and \textsc
{Lusis, A. J.} (2005). An integrative genomics approach to infer causal
associations between gene expression and disease. \textit{Nature
Genetics} \textbf{37} 710--717.

\bibitem[\protect\citeauthoryear{}{2001}]{SeCh2001}
\textsc{Sen, S.} and \textsc{Churchill, G. A.} (2001). A statistical
framework for quantitative trait mapping. \textit{Genetics} \textbf
{159} 371--387.

\bibitem[\protect\citeauthoryear{}{2000}]{SpGlSc2000}
\textsc{Spirtes, P., Glymour, C.} and \textsc{Scheines, R.} (2000).
\textit{Causation, Prediction and Search}, 2nd ed. MIT Press, Cambridge, MA.
\MR{1815675}

\bibitem[\protect\citeauthoryear{}{1990}]{VwPe1990}
\textsc{Verma, T.} and \textsc{Pearl, J.} (1990). Equivalence and
synthesis of causal models. In \textit{Readings in Uncertain Reasoning} (G. Shafer and J. Pearl, eds.).
Kaufmann, Boston.
\MR{1091985}

\bibitem[\protect\citeauthoryear{}{2006}]{WaYeScWaDrLu}
\textsc{Wang, S., Yehya, N., Schadt, E. E., Wang, H., Drake, T. A.}
and \textsc{Lusis, A. J.} (2006).
Genetic and genomic analysis of a fat mass trait with complex
inheritance reveals marked sex specificity. \textit{PLoS Genetics}
\textbf{2} e15.

\bibitem[\protect\citeauthoryear{}{2009}]{WiWiKaMiirkt2009}
\textsc{Winrow, C. J., Williams, D. L., Kasarskis, A., Millstein, J.,
Laposky, A. D., Yang, H.~S., Mrazek, K., Zhou, L., Owens, J. R.,
Radzicki, D., Preuss, F., Schadt, E.~E., Shimomura, K., Vitaterna, M.
H., Zhang, C., Koblan, K. S., Renger, J. J.} and \textsc{Turek, F. W.}
(2009). Uncovering the genetic landscape for multiple sleep-wake
traits. \textit{PLoS ONE} \textbf{4} e5161.

\bibitem[\protect\citeauthoryear{}{1934}]{Wr1934}
\textsc{Wright, S.} (1934). The method of path coefficients. \textit
{Ann. Math. Statist.} \textbf{5} 161--215.

\bibitem[\protect\citeauthoryear{}{2005}]{ZeWaZo2005}
\textsc{Zeng, Z. B., Wang, T.} and \textsc{Zou, W.} (2005). Modeling
quantitative trait loci and interpretation of models. \textit
{Genetics} \textbf{169} 1711--1725.

\bibitem[\protect\citeauthoryear{}{2008}]{ZhZhSmDrirkt2008}
\textsc{Zhu, J., Zhang, B., Smith, E. N., Drees, B., Brem, R. B.,
Kruglyak, L., Bumgarner, R. E.} and \textsc{Schadt, E. E.} (2008).
Integrating large-scale functional genomic data to dissect the
complexity of yeast regulatory networks. \textit{Nature Genetics}
\textbf{40} 854--861.

\end{thebibliography}
\end{document}